\documentclass{moriond}

\def\Journal#1#2#3#4{{#1} {\bf #2}, #3 (#4)}

\def\PRD{{\em Phys. Rev.} D}

\def\be{\begin{equation}}
\def\ee{\end{equation}}
\def\bea{\begin{eqnarray}}
\def\eea{\end{eqnarray}}

\newcommand{\mell}{\ensuremath{m_{4\ell}}}
\newcommand{\mZZ}{\ensuremath{m_{\mathrm{ZZ}}}}
\newcommand{\mH}{\ensuremath{m_{\mathrm{H}}}}
\newcommand{\GH}{\ensuremath{\Gamma_{\mathrm{H}}}}
\newcommand{\GHs}{\ensuremath{\Gamma_{\mathrm{H}}^{\mathrm{SM}}}}
\newcommand{\GHratio}{\ensuremath{\GH/\GHs}}

\newcommand{\Zo}{\ensuremath{\mathrm{Z}}}%

\newcommand{\Et}{\ensuremath{E_\mathrm{T}}}
\newcommand{\met}{\ensuremath{\Et^{\mathrm{miss}}}}
\newcommand{\MET}{\ensuremath{\Et^{\mathrm{miss}}}}
\newcommand{\mt}{\ensuremath{m_\mathrm{T}}}

\newcommand{\usedLumiB}{19.7~fb$^{-1}$}

\newcommand{\Photo}{}

\begin{document}
\vspace*{4cm}
\title{CONSTRAINTS ON THE HIGGS-BOSON TOTAL WIDTH USING $\mathrm{H}^{*}(126) \rightarrow \mathrm{ZZ}$ EVENTS}

\author{ R. COVARELLI, for the CMS collaboration }

\address{Department of Physics and Astronomy, University of Rochester, 14627, Rochester, NY, United States}

\maketitle\abstracts{
Constraints are set on the Higgs boson decay width, $\Gamma_{\mathrm{H}}$, using off-shell
production and decay to ZZ in the four-lepton ($4\ell$), or two-lepton-two-neutrino ($2\ell 2\nu$) final states.
The analysis is based on the data collected in 2012 by the
CMS experiment at the LHC, corresponding to an integrated luminosity 
${\cal L} = 19.7\, \mathrm{fb^{-1}}$ at 
$\sqrt{s} = 8  \, {\rm TeV}$. A maximum-likelihood fit of invariant mass 
and kinematic discriminant distributions in the $4\ell$ case and of 
transverse mass or missing energy distributions in the $2\ell2\nu$ case 
is performed.
The result of it, combined with the 
$4\ell$ measurement near the resonance peak, leads to an upper limit 
on the Higgs boson width
of $\Gamma_{\mathrm{H}} < 4.2 \times \Gamma_{\mathrm{H}}^{\mathrm{SM}}$ at the 95\% confidence level, assuming 
$\Gamma_{\mathrm{H}}^{\mathrm{SM}} = 4.15 \, {\rm MeV}$.
}

\section{Introduction}

After the discovery of a particle consistent with the standard 
model (SM) Higgs boson, direct constraints on the new boson width (\GH)
of 3.4 GeV at the 95\% confidence level (CL) have been reported by the CMS experiment in the 
$4\ell$ decay channel~\cite{Chatrchyan:2013legacy}.
With the current data, the sensitivity for a direct width measurement 
at the resonance peak 
is therefore far from the SM Higgs boson expected width of around 4 MeV. 

It has been proposed~\cite{CaolaMelnikov:1307.4935}
to constrain the Higgs boson width using the off-shell production
and decay in ZZ, since, in the gluon-gluon fusion production mode, the off-shell 
production cross section has been shown to be sizable at high
ZZ invariant mass (\mZZ)~\cite{Passarino:2012ri,Kauer:2012hd}.

The production cross section
as a function of \mZZ\ can be written as:
\begin{equation}
\label{eq:fava}
\frac{d\sigma_{\mathrm{gg} \rightarrow \mathrm{H} \rightarrow \mathrm{ZZ}}}{d\mZZ^2}  
\propto g^2_{\mathrm{ggH}}g^2_{\mathrm{HZZ}}
\frac{F(\mZZ)}{(\mZZ^2 - \mH^2)^2 + \mH^2\GH^2}\, , 
\end{equation}
where $g_{\mathrm{HZZ}}$ ($g_{\mathrm{ggH}}$) represent the (effective) couplings of the Higgs boson to Z bosons (gluons), $m_H$ is
the measured Higgs pole mass, 
and $F(\mZZ)$ is a function which depends on the Higgs boson production and decay amplitudes.
In the on-shell (off-shell) regions, the integrated (differential) cross sections are respectively:
\begin{equation}
\sigma^{\mathrm{on-shell}}_{\mathrm{gg} \rightarrow \mathrm{H} \rightarrow \mathrm{ZZ}} = 
\frac{\kappa_{\mathrm{g}}^2\kappa_{\Zo}^2}{r} (\sigma\cdot\mathrm{\cal B})_{\mathrm{SM}}
\equiv \mu (\sigma\cdot\mathrm{\cal B})_{\mathrm{SM}},
\label{eq:resonnantregion}
\end{equation}
and:
\begin{equation}
\frac{d\sigma^{\mathrm{off-shell}}_{\mathrm{gg} \rightarrow \mathrm{H} \rightarrow \mathrm{ZZ}}}{d\mZZ} = 
\kappa_{\mathrm{g}}^2\kappa_{\Zo}^2 \cdot \frac{d\sigma^{\mathrm{off-shell, SM}}_{\mathrm{gg} \rightarrow \mathrm{H} \rightarrow \mathrm{ZZ}}}{d\mZZ}
= \mu r \frac{d\sigma^{\mathrm{off-shell, SM}}_{\mathrm{gg} \rightarrow \mathrm{H} \rightarrow \mathrm{ZZ}}}{d\mZZ},
\label{eq:crosssect}
\end{equation}
where $(\sigma\cdot\mathrm{\cal B})$ is the cross section times branching fraction to ZZ, all quantities are expressed as adimensonal ratios to their SM values ($\kappa_g = g_{\mathrm{ggH}} / g_{\mathrm{ggH}}^{\mathrm{SM}}$, $\kappa_Z=g_{\mathrm{HZZ}}/g_{\mathrm{HZZ}}^{\mathrm{SM}}$, $r = \GHratio$), and the quantity $\mu$ defined by this
relationship is referred to as the ``signal strength''.
From Eqs.~(\ref{eq:resonnantregion}, \ref{eq:crosssect}) it is clear that the ratio of off-shell and on-shell production and decay rates
in the $\mathrm{H}\rightarrow\mathrm{ZZ}$ channel leads to a direct measurement of $\GH$ as long as the
ratio of coupling constants remains invariant at the low and high $\mZZ$ values. A similar formalism can be used for the vector boson fusion (VBF) production. 

We obtain an upper bound on \GH~from the comparison of off-shell production and decay distribution in the 
$\mathrm{H}\rightarrow\mathrm{ZZ}\rightarrow 4\ell$ and $\mathrm{H}\rightarrow\mathrm{ZZ}\rightarrow2\ell2\nu$ channels,
and the $4\ell$ on-shell rate, where $\ell = \mathrm{e},\mu$. 
The analysis is based on the dataset collected by the CMS experiment during the 2012 LHC running period, which corresponds to an integrated luminosity of \usedLumiB~of pp
collisions at a center-of-mass energy of $\sqrt{s}$ = 8 TeV. Details of this analysis can be found in~\cite{nostraPAS,nostraWiki}. 
A detailed description of the CMS detector can be found in Ref.~\cite{CMSDETECTOR}.
Concerning lepton and missing transverse energy reconstruction and event selection, this analysis uses the same algorithms  
as in Refs.~\cite{Chatrchyan:2013legacy,2l2nu}. 
 
\section{Analysis strategy}

As shown above, once a value of $\mu$ is constrained from an independent measurement or calculation, the off-shell cross section as a function of \mZZ\ is proportional to \GH. We use two different assumptions for $\mu$, i.e. the measured value from the the $4\ell$ on-shell analysis~\cite{Chatrchyan:2013legacy}, or $\mu=1$ assuming the SM expectations in the peak, with the expected uncertainties from the same analysis.

The VBF mechanism also leads to significant off-shell Higgs production.
The signal strengths can be considered separately for the gluon-gluon 
fusion ($\mu_{\mathrm{F}}$) and VBF ($\mu_{\mathrm{V}}$) production mechanisms.
However, because of the limited precision obtained on these quantities from
the current data, we assume in this analysis that 
$\mu_{\mathrm{V}}=\mu_{\mathrm{F}}=\mu$.

At large \mZZ, interference between signal and background for the processes with the same initial and final states is not negligible and must be taken into
account. Therefore the event likelihood can be written as:
\begin{eqnarray}
&& {\cal P}_{\rm tot}^{\rm off\mbox{-}shell}(\vec{x}) = 
\left[ 
\mu r \times {\cal P}^{\rm gg}_{\rm sig}(\vec{x}) +  \sqrt{\mu r} \times  {\cal P}^{\rm gg}_{\rm int}(\vec{x}) + {\cal P}^{\rm gg}_{\rm bkg}(\vec{x})
\right] +
\nonumber \\
 && + 
\left[
\mu r \times {\cal P}^{\rm VBF}_{\rm sig}(\vec{x}) +  \sqrt{\mu r } \times  {\cal P}^{\rm VBF}_{\rm int}(\vec{x}) + {\cal P}^{\rm VBF}_{\rm bkg}(\vec{x})
\right]  + 
~{\cal P}^{\mathrm{q}\bar{\mathrm{q}}}_{\rm bkg}(\vec{x}) + \mathrm{(other~backgrounds)}
\label{eq:pdf-prob-vbf}
\end{eqnarray}
where ${\cal P}_{\rm sig}$, ${\cal P}_{\mathrm{int}}$ and ${\cal P}_{\rm bkg}$ are signal,
interference, and background probability functions, respectively, for gluon-gluon fusion and VBF production,
and defined as functions of the variables used in each analysis.

\section{Monte Carlo simulation}

The Monte Carlo (MC) samples used in this analysis
are the same as those described in Refs.~\cite{Chatrchyan:2013legacy} and~\cite{2l2nu}. 

Additionally, $\mathrm{gg} \rightarrow 4\ell$ ($2\ell 2\nu$) events,
as well as $\mathrm{qq'} \rightarrow\mathrm{ZZ} \mathrm{qq'} \rightarrow 4\ell\mathrm{qq'}$ ($2\ell 2\nu \mathrm{qq'}$) VBF events, have been 
generated at the leading order (LO) including the Higgs signal
as well as the background and interference using 
different MC generators: \textsc{gg2VV} 
3.1.5~\cite{Kauer:2012hd,Kauer:2012ma}, \textsc{MCFM} 
6.7~\cite{CampbellEllisWilliams:1311.3589v1,MCFM}, and \textsc{PHANTOM}~\cite{Ballestrero:2007}. Samples have been generated with MSTW2008 LO
parton density functions (PDFs) and the renormalization and factorization scales are proportional to \mZZ~(``running'' scales). 

We apply to next-to-next-to-leading order (NNLO) corrections (``K-factors'') 
as a function of $\mZZ$~\cite{Passarino:1312.2397v1} to the gluon-fusion signal process and, even
if exact calculations of the background process are limited to LO, we assign 
the same K-factor to it, relying on soft-collinear approximation~\cite{Bonvini:1304.3053}. For VBF production, the event yield is normalized to the cross section at NNLO~\cite{LHCHiggsCrossSectionWorkingGroup:2011ti},
with a normalization factor independent of $m_{4\ell}$.

\section{$\mathrm{H}\rightarrow\mathrm{ZZ}\rightarrow 4\ell$ analysis}
\label{sec:quattrol}

\begin{figure}[h]
\begin{center}
\includegraphics[width=0.3\textwidth]{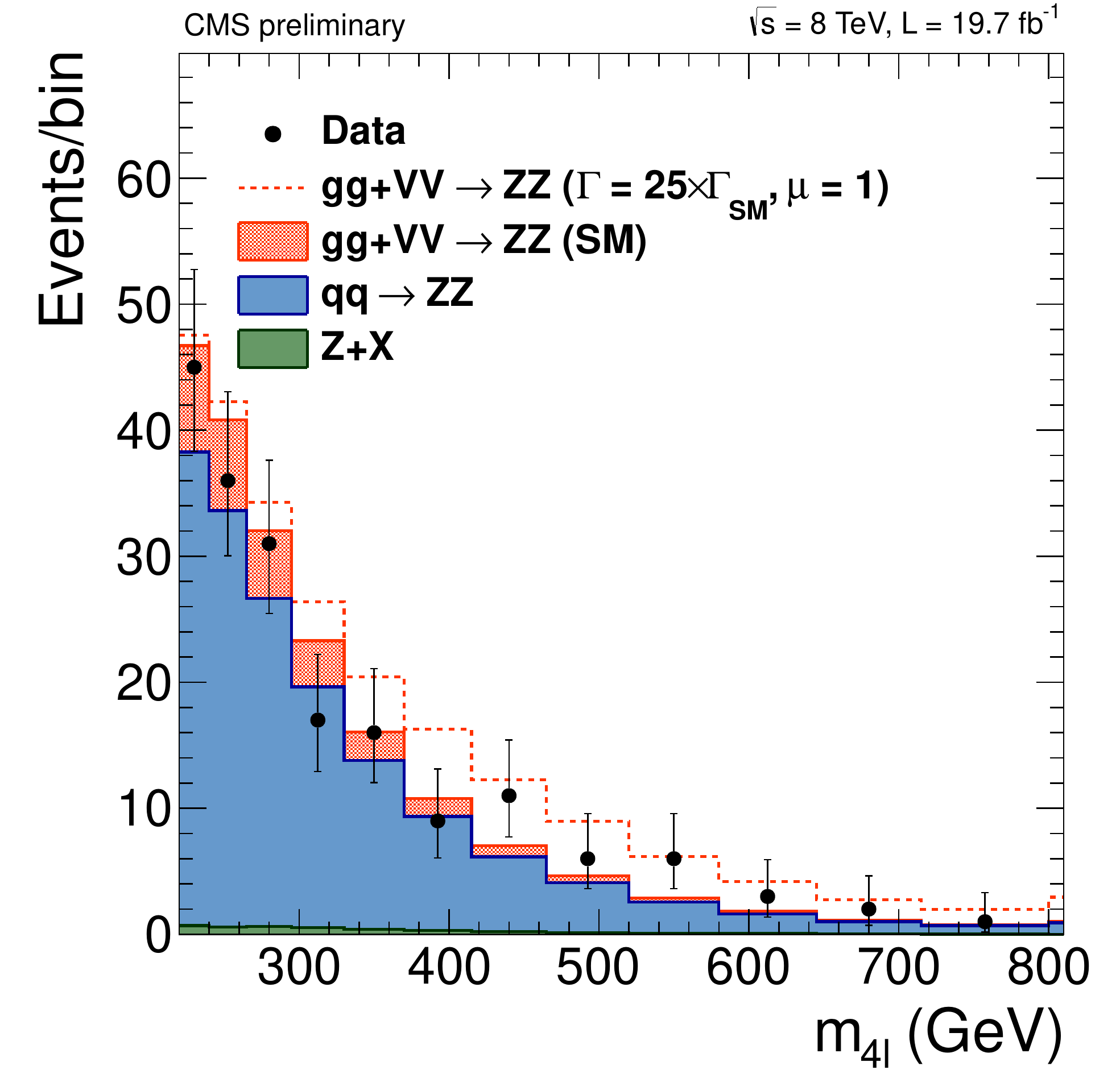}
\includegraphics[width=0.3\textwidth]{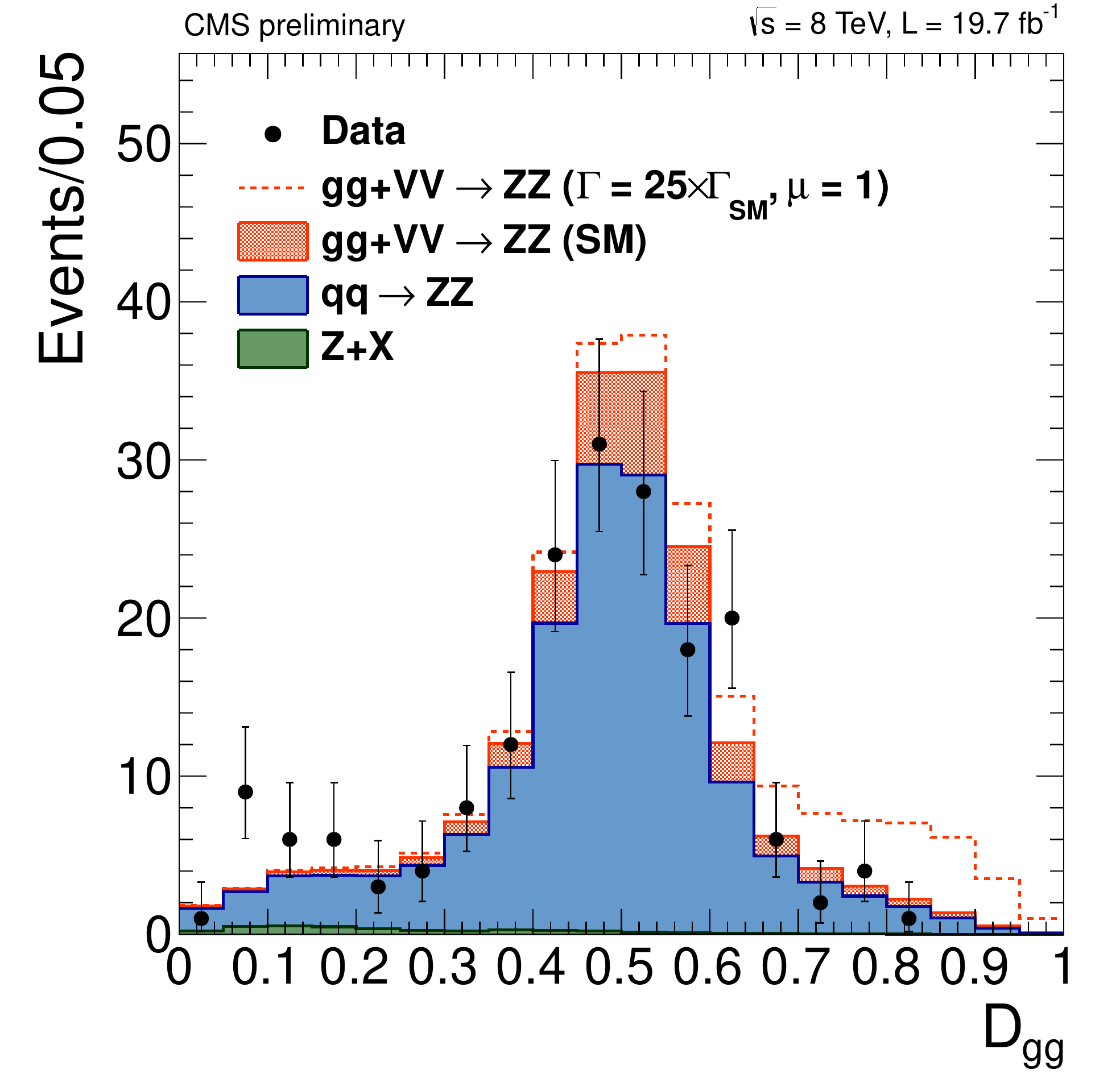}
\includegraphics[width=0.32\textwidth]{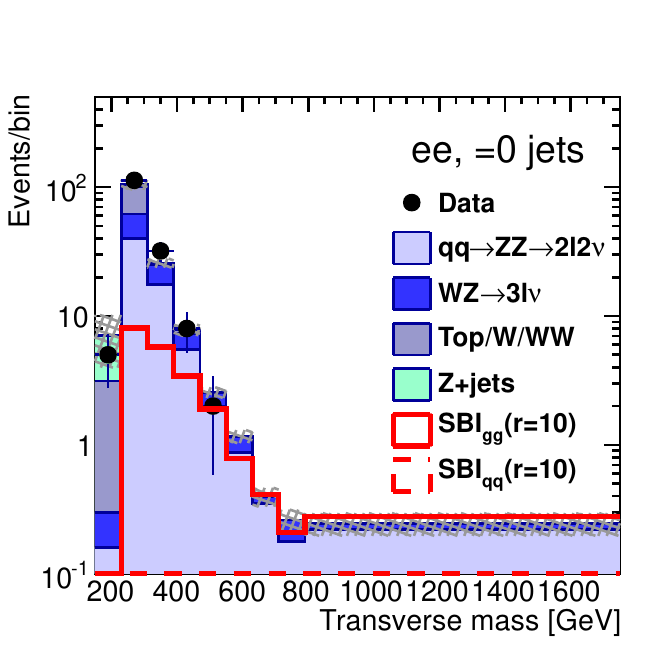}
\caption{Distributions of the discriminating variables used in the $4\ell$
analysis: $\mell$ (left) and ${\cal D}_{\mathrm{gg}}$ (center) for the data and all the expected contributions.
The latter are shown for the SM expectation, as well as for an hypothesis 
corresponding to $r=25$, and include both the gg and the VBF processes.
Distribution of the $\mt$ variable (right) in the in the $2\mathrm{e}2\nu$
0-jet category, for the data and the expected contributions. The shapes of
gg and VBF inclusive processeses (SBI = signal, background and interference)
for a $r=10$ scenario are superimposed.}
\label{fig:mss-discr}
\end{center}
\end{figure}

In addition to the reconstruction, selection, and analysis methods developed in~\cite{Chatrchyan:2013legacy}, the $4\ell$ 
off-shell analysis uses a dedicated kinematic discriminant ${\cal D}_{\mathrm{gg}} $ which
describes the production and decay dynamics in the ZZ center-of-mass frame using as observables the two dilepton masses as well as five independent angles~\cite{Gao:2010qx}.
The discriminant is defined as ${\cal D}_{\mathrm{gg},a} \equiv {\cal P}_{\mathrm{gg},a} / ({\cal P}_{\mathrm{gg},a}  + {\cal P}_{\mathrm{q} \bar{\mathrm{q}}})$, where ${\cal P}_i$ is the probability for a $4\ell$ event to come either from
$\mathrm{gg}\rightarrow ZZ$ or $\mathrm{q} \bar{\mathrm{q}} \rightarrow ZZ$ processes. 
The discriminant is defined for a signal-weight parameter $a$, 
where $a=1$ corresponds to the SM.
We set $a=10$ in constructing the discriminant, since an exclusion of the order of $r=10$ is expected to be achieved. 
Figure~\ref{fig:mss-discr} shows the distribution of the the $4\ell$ invariant mass (left) and the
${\cal D}_{\mathrm{gg}}$ variable (center) for all expected contributions, as well as for the data.

\section{$\mathrm{H}\rightarrow\mathrm{ZZ}\rightarrow 2\ell 2\nu$ analysis}

Following the same event reconstruction and
selection used in previous searches for high-mass Higgs bosons~\cite{2l2nu}, 
the selected events are split according to lepton flavors (e and $\mu$) and jet categorization (0 jets, $\geq$1 jet, and ``VBF-like'', i.e. two jets satisfying $m_{jj} > 400$ GeV and $\Delta\eta_{jj} > 4$).
The transverse mass ($\mt$) distributions for the 0 and $\geq$1 jets categories and the missing transverse energy ($\met$) distribution
for the VBF-like category are used as final discriminant variables.
The $\mt$ variable is defined as follows:

\begin{equation}
\mt^{2} = \Bigg[\sqrt{{p_{\mathrm{T},\ell\ell}}^2 + {m_{\ell\ell}}^2} + \sqrt{{\MET}^2 +
{m_{\ell\ell}}^2}\Bigg]^2 - \Big[\vec{p}_{\mathrm{T},\ell\ell} + \vec{E}_T^\mathrm{miss}\Big]^2
\end{equation}

where $\vec{p}_{\mathrm{T},\ell\ell}$ and $m_{\ell\ell}$
are the measured transverse momentum and invariant mass of the dilepton system, respectively.
One of the \mt~distributions is shown in Fig.~\ref{fig:mss-discr} (right).

\section{Systematic uncertainties}
\label{sec:syst}

The main systematic uncertainty in this analysis comes from the measured value
of $\mu$: in the approach using its expected (observed) value, it is taken
from Ref.~\cite{Chatrchyan:2013legacy} to be $1.00^{+0.27}_{-0.24}$ ($0.93^{+0.26}_{-0.24}$). In the approach used in this analysis, all signal systematics
for the $4\ell$ final state
depending only on normalization cancel when using the measured 
on-shell signal strength as a reference.
Other experimental systematic uncertainties are considered on the 
amount of reducible background in the $4\ell$ analysis and in the evaluation 
of \MET and the b-jet veto efficiency in the $2\ell2\nu$ analysis.

Theoretical uncertainties are 
important in this analysis for the signal and interference
contributions and for the $\mathrm{q} \bar{\mathrm{q}} \rightarrow ZZ$ background. 
QCD renormalization and factorization scales are varied by a
factor two both up and down, and uncertainties from PDF
variations are extracted by changing PDF sets. These uncertainties are computed
on LO MC and K-factors and applied consistently.
To account for the limited knowledge on the $\mathrm{gg} \rightarrow \Zo\Zo$ continuum background cross
section at NNLO (and therefore on the interference), we assign an additional systematic uncertainty of 10\%. 

\section{Results}

\begin{figure}[hbtp]
\begin{center}
\includegraphics[width=0.32\textwidth]{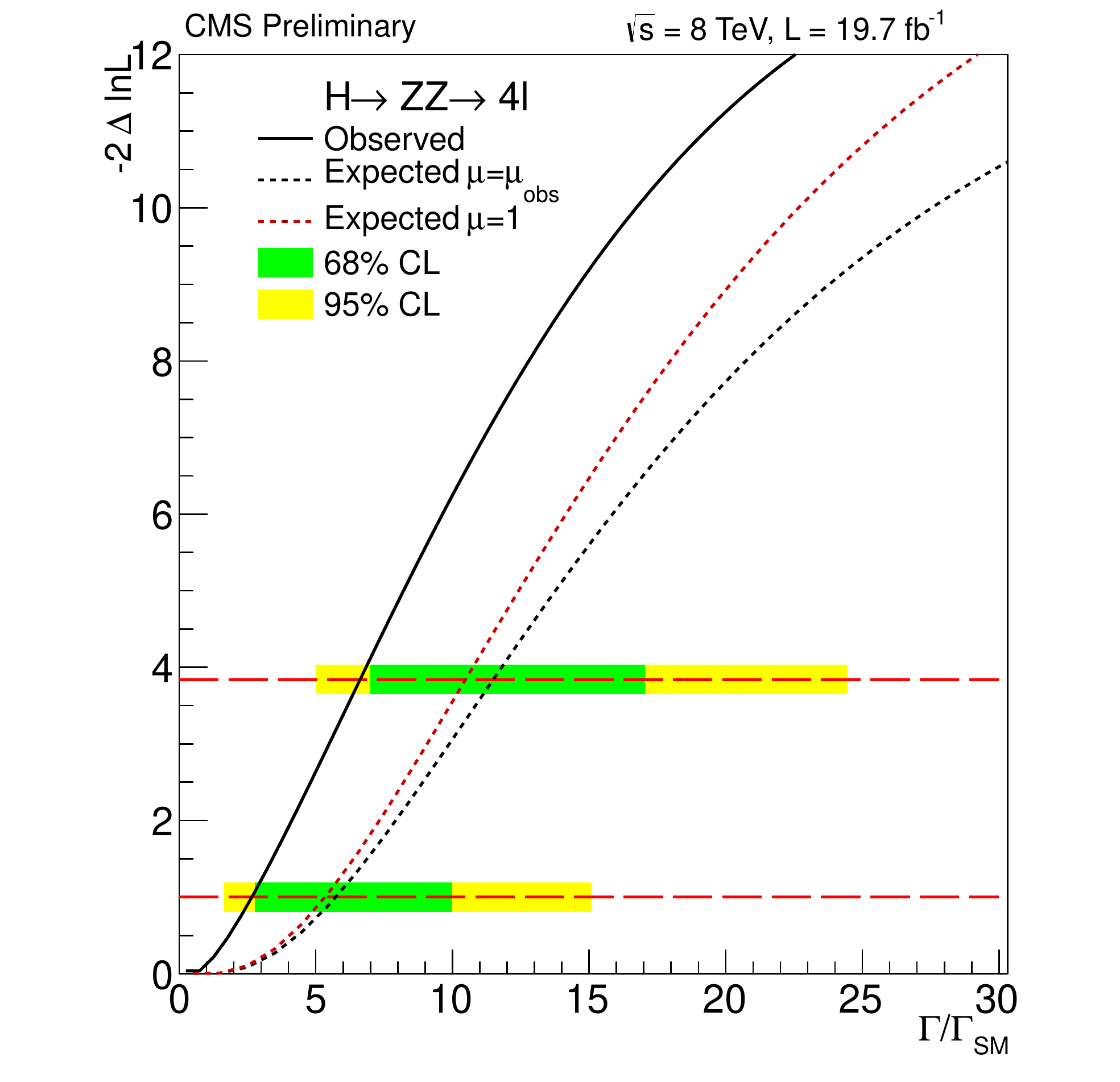}
\includegraphics[width=0.32\textwidth]{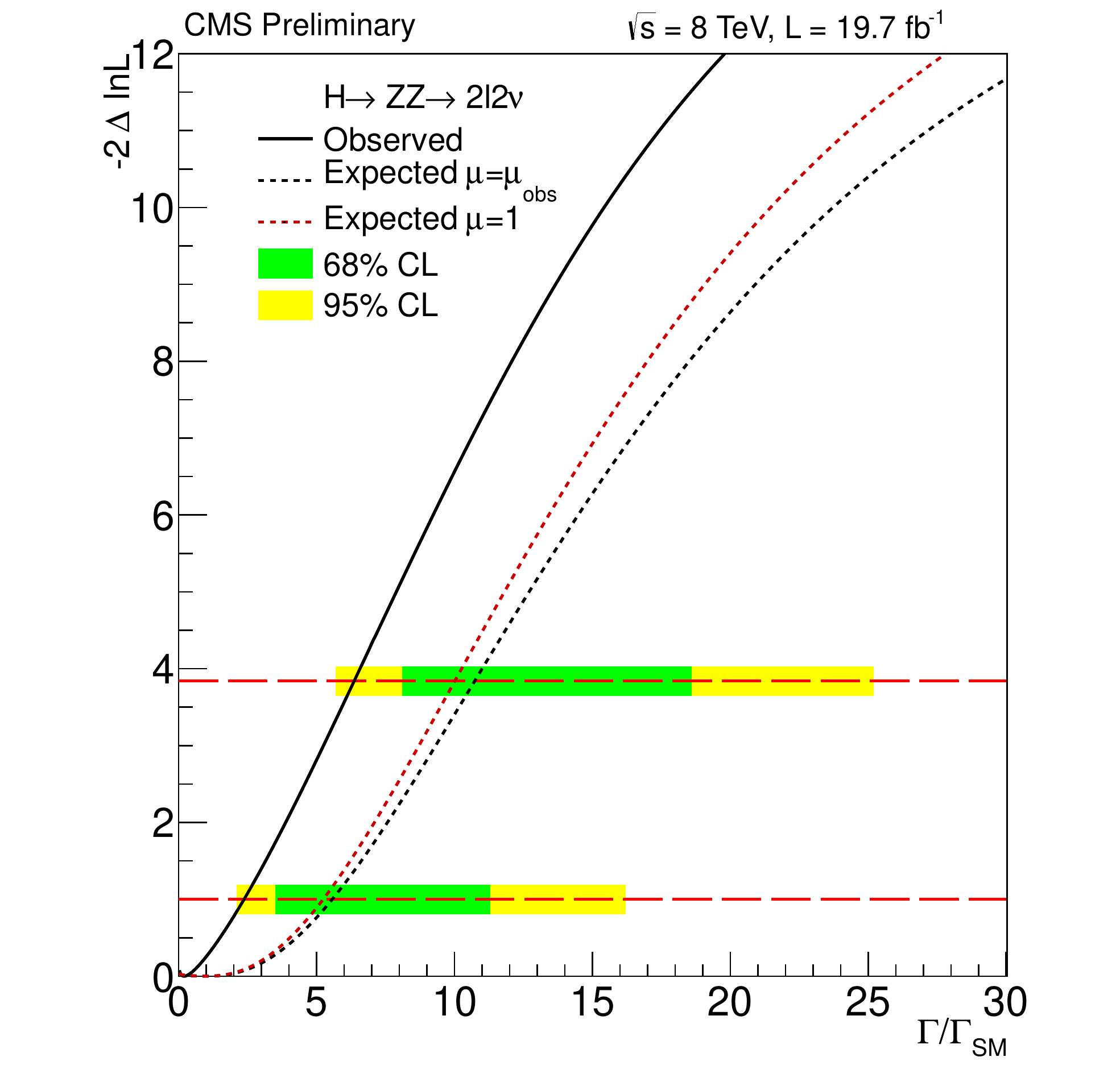} 
\includegraphics[width=0.32\textwidth]{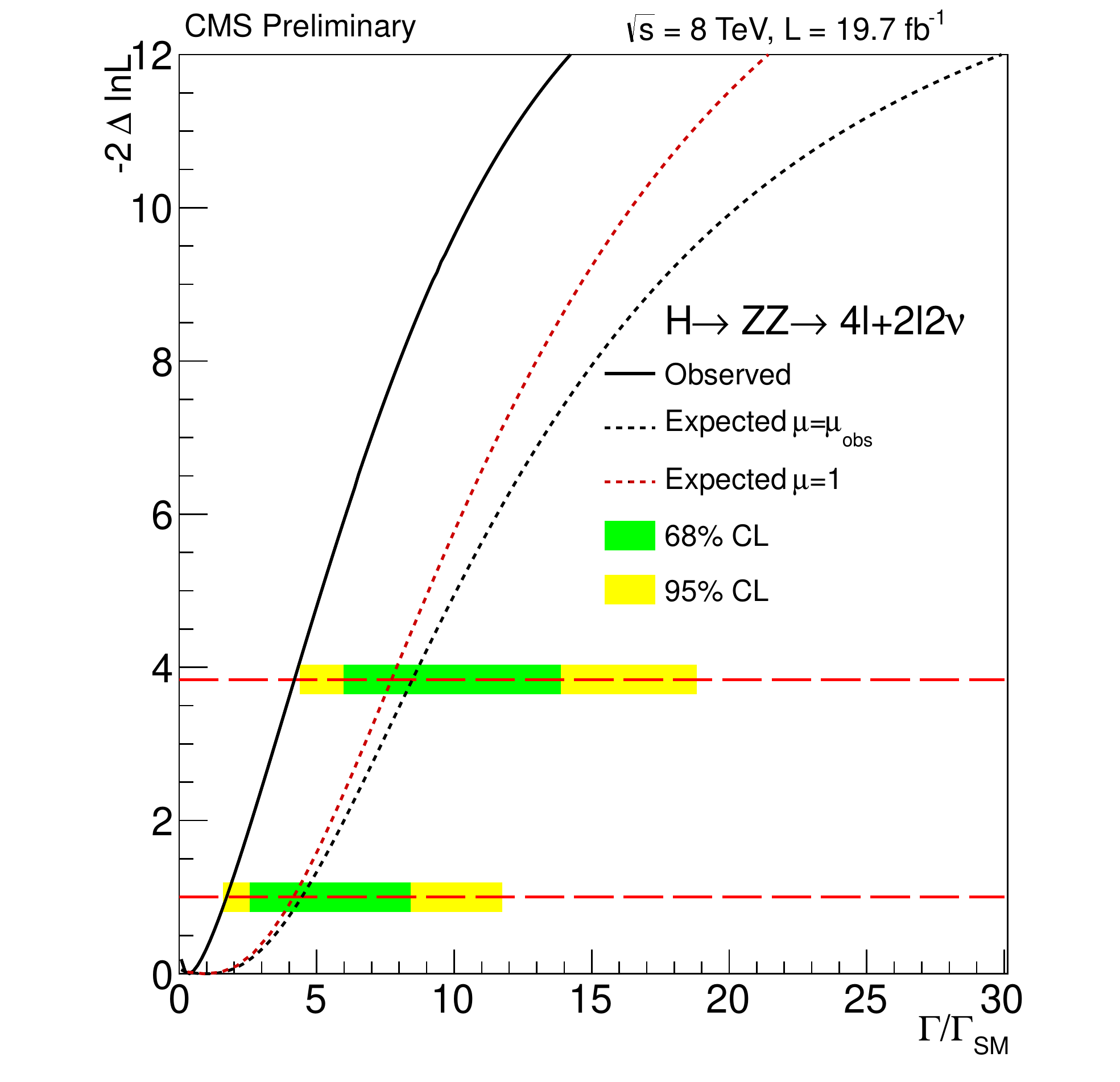}
\caption{
Negative log-likelihood scans as a function of $r=\GHratio$ for the
$4\ell$ (left) and $2\ell2\nu$ (center) analyses separately and for the combined 
result (right). Green and yellow
bands correspond respectively to 68\% and 95\% quantiles of the 
distribution of the negative log-likelihood at the corresponding 
value on the red line, using MC pseudo-experiments.  
}  
\label{fig:fit2Dcombine}
\end{center}
\end{figure}

Using the normalization and shape of the signal and background distributions, 
which are derived either from MC or control samples, 
an unbinned
maximum-likelihood fit of the data is performed, 
where systematic uncertainties are included as nuisance parameters.
In the $4\ell$ analysis the kinematic discriminant is combined with $\mell$ in a two-dimensional fit, while $\mt$ or $\met$ distributions 
are used for the $2l2\nu$ channel.
Fit results are shown in Fig.~\ref{fig:fit2Dcombine}, in the form of  
negative log-likelihood scans as a function of $r$. The red 
dashed lines correspond to 68\% and 95\% CL exclusion values. 
Table~\ref{tab:finalres} shows the results obtained using the 
observed value of $\mu$.
Combination of the two channels results in an observed (expected) exclusion of
 $\Gamma_{\mathrm{H}} \le 4.2~(8.5) \times \Gamma_{\mathrm{H}}^{\mathrm{SM}}$ at the 95\% CL, or $\Gamma_{\mathrm{H}} \le 17.4~(35.3)$ MeV.

\begin{table}[hbtp]
\begin{center}
\footnotesize
\begin{tabular}{l|c|c|c}
\hline
 & $4\ell$ & $2\ell2\nu$ & Combined \\
\hline
Expected 95\% CL limit, $r$ & 11.5 & 10.7 & 8.5\\
Observed 95\% CL limit, $r$ & 6.6 & 6.4 & 4.2 \\
Observed 95\% CL limit, $\GH$ (MeV) & 27.4 & 26.6 & 17.4 \\
\hline
Observed best fit, $r$ & 0.5 $^{+2.3}_{-0.5}$ & 0.2 $^{+2.2}_{-0.2}$ & 0.3 $^{+1.5}_{-0.3}$ \\
Observed best fit, $\GH$ (MeV) & 2.0 $^{+9.6}_{-2.0}$ & 0.8 $^{+9.1}_{-0.8}$ & 1.4 $^{+6.1}_{-1.4}$ \\
\hline
\end{tabular}
\end{center}
\caption{Expected and observed 95\% CL limits and fitted values 
of $r$ or \GH~for 
the $4\ell$ analysis, the $2\ell2\nu$ analysis and for the combination, using the observed value of $\mu$.
}
\label{tab:finalres}
\end{table}

\end{document}